\documentclass[3p, 12pt, onecolumn]{elsarticle}
\usepackage[T1]{fontenc}
\usepackage{amssymb}
\usepackage{caption}
\usepackage{subcaption}
\usepackage{graphicx,wrapfig,lipsum}
\usepackage{amscd,amsmath,verbatim}
\usepackage{amsfonts,epsfig}
\usepackage{mathptmx}
\usepackage{setspace}
\usepackage{epsfig}
\usepackage{url}
\usepackage{color}

\pdfoutput=1

\begin{document}

\journal{Chaos, Solitons \& Fractals}

\begin{frontmatter}

\title{Lattice Gas model to describe a nightclub dynamics}

\author{Eduardo Velasco Stock, Roberto da Silva} 

\address{Instituto de F\'{i}sica, Universidade Federal do Rio Grande do Sul,
Porto Alegre Rio Grande do Sul, Brazil
}

\begin{abstract}
In this work, we propose a simple stochastic agent-based model to describe the revenue dynamics of a nightclub venue based on the relationship between profit and spatial occupation. The system consists of an underlying square lattice (nightclub's dance floor) where every attendee (agent) is allowed to move to its first neighboring cells. Each guess has a characteristic delayed time between drinks, denoted as $\tau$, after which it will show an urge to drink. At this moment, the attendee will tend to move towards the bar where a drink will be bought. After it has left the bar zone, $\tau$ time steps should pass so it shows once again the need to drink. Our model among other points show that it is no use filling the bar to obtain profit, and optimization should be analyzed. This can be done in a more secure way taking into consideration the ratio between income and ticket cost. 
\end{abstract}

\end{frontmatter}

\section{Introduction}

\label{Section:Introduction}

Is greed a problem? Yes, when greed culminates in a tragedy. At the head of
many bars or nightclub owners, when more overcrowded better the business.

However, it is not entirely true. In the early hours of the 27th of January,
2013, a fire broke loose at the Kiss nightclub in Santa Maria, Brazil. It
started due to the negligent use of firecrackers indoors by the music band
scheduled to play that night. In the following events, 242 people lost their
lives, and 600 were left injured \cite{dalponte2015}.

In tragic events like this, several factors can increase the death toll,
such as the absence of proper firefighting equipment or personnel, panic
stamped, drug or alcohol abuse \cite%
{Clapp2017,https://doi.org/10.1080/09595230802089719}, poorly designed
evacuation routes, but mainly due to overcrowding.

The survivors of the Kiss nightclub fire have shown severe psychological and
physical sequelae \cite{Calegaro2019}, as well as The Station nightclub
patrons \cite{10.1371/journal.pone.0115013,10.1371/journal.pone.0047339}, in
a similar nightclub fire \cite{station_fire_report_2005} on February the
20th, 2003.

The severity of tragedies in mass gathering events such as the Hillsborough
Stadium tragedy (1989), the Love Parade disaster in Germany (2010), and the
frequent Indian religious tragedy with an overwhelmingly large number of
attendees has called the attention of physicists.

In this direction, they elaborate reliable mathematical models \cite%
{Helbing-simulating-dynamics-features-escape-panic-2000}, experimental
studies \cite{STICCO2021125299}, and even video analysis of such scenarios 
\cite{Loveparade-disaster-video-analysis-2012}.

Back in 1995, Helbing and Moln\'{a}r proposed a pioneer "social force" model 
\cite{Helbing-social-force-model-pedestrian-dynamics-1995} to describe the
counterflowing pedestrian dynamics and crossing through bottlenecks.

Since then, different applications and variations of the social force model
have successfully captured some observed phenomena in the context of
pedestrian dynamics. The wide variety of phenomena can comprehend
stop-and-go waves \cite%
{Andrea-modeling-stop-and-go-wave-2010,Helbing-dynamics-of-crowd-disasters-empirical-study-2007}%
, arch formation in bottlenecks \cite{PhysRevLett.97.168001,Bottinelli2016},
vortex emergence from Mosh pits in Metal concerts \cite%
{PhysRevLett.110.228701}, jamming \cite{PhysRevE.92.042809} and lane
formation \cite%
{pedestrian_flow_cellular_automata2015,Helbing-social-force-model-pedestrian-dynamics-1995,PhysRevE.85.066128}%
, just to cite a few.

A different approach for pedestrian dynamics modeling is considering the
dynamics on a lattice and using mathematical tools from statistical
mechanics. In this context, cellular automaton models \cite%
{BURSTEDDE2001507,friction_clogging_automata_pedestrian_Ansgar2003} are an
excellent way to describe short-range interactions of many agents on the
lattice. In \cite{Stock_JSTAT_2019,nosso_PRE_clogging_2019}, we proposed an
agent-based model of counterflowing particles on a square lattice as an
extension of a previous model to describe a one-dimensional system of
counterflowing streams \cite{Roberto-pedestrian-dynamics-2015}. In that
extension, we could reproduce the emergence of jamming, lane formation, and
phase coexistence.

An essential aspect of people's evacuation from enclosed environments is the
competitive behavior that emerges when they have to enter a passage with a
specific geometric layout.

From that perspective, one observes important phenomena, such as the
power-law tail of time distribution between consecutive individuals crossing
narrow doors \cite{Hidalgo2017} or the transition from the normal to the
power-law distribution of crossing times of rectangular corridors with
obstacles \cite{mestrado_eduardo_roberto2017}.

Additionally, one observes other exciting effects such as "faster-is-slower"
on the distribution of evacuation times \cite{PhysRevE.94.022313,Pastor2015}
or boarding airplane queues \cite{Erland2021}, or even self-organized
behavior due to asymmetries of the sidewalls of corridors \cite%
{Andre-keep-left-behavior-2016}.

Even though many theoretical and experimental works consider describing the
phenomena emergent from pedestrian dynamics to predict and avoid tragedies,
there are almost no studies concerning the relationship between the revenue
of nightclubs and the occupation of such places. More precisely, the study
of universal aspects from the point of view of Physics.

These considerations bring us back to the fire in the Kiss nightclub in
Brazil. After almost nine years after this tragedy, the case has gone to a
jury trial where several witnesses testified that the nightclub was at full
capacity at the tragic venue, which was considered by the prosecutors as a
wanted scenario by the nightclub owners with the intent of maximizing
profits.

Paradoxically, one of the defendants (nightclub owners) stated that full
capacity was not interesting for the business since it does not translate
into consumption at the nightclub bar. From the point of view of ticket
sales, it is straightforward that revenue grows linearly with the density of
attendees once each attendee pays once to enter the nightclub (if attendance
has a cost).

However, from the point of view of bar revenue, an important question is:
how the pedestrian dynamics affect the consumption of beverages inside the
nightclub? Also, what is the best ratio between ticket and drink prices to
maximize a nightclub's revenue, while keeping the place safe? Is it in the
interest of the nightclub owner to have its nightclub at full capacity?

To answer these questions, we propose in this work a simple agent-based
model based on a driven-diffusive lattice gas dynamics \cite%
{original_lattice_gas-Katz} to describe nightclub-like dynamics and to
examine the intricated relationship between monetary aspects and the
nightclub occupation.

Our model consists of an adaptation of a stochastic model of particles
proposed initially by Montrol in \cite{Montroll-stochastic-processes-1979},
where we define a driven-diffusive lattice gas dynamics with a static floor
field \cite%
{friction_clogging_automata_pedestrian_Ansgar2003,PhysRevE.94.022313} to
describe the motion of attendees getting to and leaving the nightclub bar.
To capture realistic aspects of attendees' drinking habit, we implemented
the concept of memory to describe the time it takes for a person to show an
urge to have another drink. This concept has already been used in game
theory models for decision-making when switching strategies \cite%
{Burridge2015}.

We organize the paper in the following way. In section \ref{Sec:Model}, we
present the model, including equations and rules that govern the dynamics of
our lattice gas. In the sequence, we explore the results of this modeling by
showing how the revenue of nightclub-like enclosures and occupations are
related, considering the effects of bias and distributions on the
consumption frequency. Finally, we summarize our results and present some
conclusions in the section \ref{Sec:Conclcusions}.

\section{Model}

\label{Sec:Model}

\begin{figure*}[t]
\begin{center}
\includegraphics[width=\textwidth]{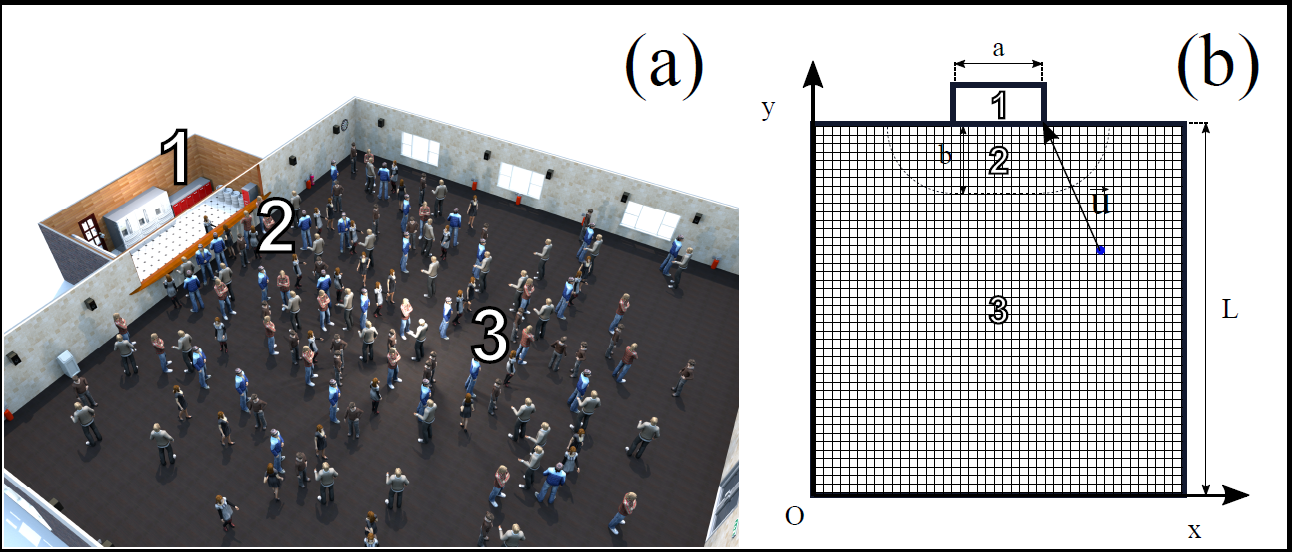}
\end{center}
\caption{(a) A depiction of a nightclub with a simple layout. (b) Lattice
representation of situation (a). Region one corresponds to the bar. Regions
two and three effectively form the lattice at which agents can move around.
Region two is an area of turbulence where agents come and go due to the bar.}
\label{Fig:night_club}
\end{figure*}

One can imagine a nightclub pictorially as a dance floor with a simple
layout represented by Fig. \ref{Fig:night_club} (a). In this plot, we have
three important regions: Region one corresponds to the bar. Region two is
the surroundings of the bar. 
%, where people walk with their drinks in the direction
%of different bar positions.
Finally, region 3 is where the people are furthest from the bar.

We define our model as a system of a fixed number of agents (attendees)
denoted by $N$, where the agents move on an underlying square lattice
composed of $L^{2}$ cells of unitary size, which is a translation of the
scenario considered in Fig. \ref{Fig:night_club} (a), here represented by
Fig. \ref{Fig:night_club} (b).

For clarity, the bar comprehends a region outside the lattice boundaries
(inaccessible to agents) as depicted in Figs. \ref{Fig:night_club} (a) and
(b). Thus, for buying a drink, agents must reach any lattice cell contiguous
to the bar of length $a$ (for construction $a\leq L$), i.e., any cell at the
position $(x,y)=(\frac{L-a}{2},L)$ to $(x,y)=(\frac{L+a}{2},L)$. 
%When an agent reaches the bar, it will remain in that place for $t_{w}$ time steps, which is the average time customers wait for a drink.

After an attendee buys a drink in the bar, one computes a sale, and the
nightclub's revenue increases by the amount $c_{d}$, which is the drink
cost. It is reasonable to expect that after an agent gets a drink, it will
try to move away from the bar. It means to leave the \textit{bar zone},
which is a turbulent region (shown as region 2 in Figure \ref{Fig:night_club}
(b)) due to the intense flow of people leaving with beverages, people trying
to reach the bar and people walking around randomly.

More precisely, region two is composed of all cells that are at a distance
to the bar equal or smaller than $b$, as shown in Figure \ref{Fig:night_club}
(b).

To describe such buying and drinking dynamics, we define that each agent can
be in one of three intrinsic situations (states): thirsty $(i)$, served $%
(ii) $, or dancing $(iii)$. In state $(i)$, the agents present a biased
motion toward the bar. In state $(ii)$, agents tend to move away from the
bar zone. In state $(iii)$, agents move randomly through the lattice. The
transition between these three states occurs in a cyclic fashion of $%
(i)\rightarrow (ii)\rightarrow (iii)\rightarrow (i)$ according to the space
and time variables. For instance, a given agent in the state $(i)$
immediately switches to state $(ii)$ when it gets to a cell adjacent to the
bar. An agent in state $(ii)$ immediately changes to state $(iii)$ when it
crosses from region two to region three (shown in Figure \ref{Fig:night_club}
(a) and (b)) for the first time since it switched to state (ii). Finally, an
agent in the state $(iii)$ moves randomly throughout the lattice. It stays
in the state $(iii)$ for $\tau $ time steps before switching to state $(i)$.

The parameter $\tau $ is related to memory time and it is associated to each
subject's drinking characteristics in the sense that an agent will only
present the urge to buy its next drink after $\tau $ time steps have passed
since it left region 2.

By defining $t_{\max }$ as the nightclub venue duration, one considers that
a given attendee with a memory $\tau >t_{\max }$ can be said to be
abstemious. In contrast, another given attendee with low $\tau <t_{\max }$
represents a drinking habit ranging from a social drinker to a compulsive
drinker.

By occupying any one of these three possible states, a given attendee $k$
after $l$ time steps can move to one of its first neighboring cells
according to the following transition probabilities defined by our lattice
gas: 
\begin{equation}
\begin{array}{ccc}
P_{(x,y)\rightarrow (x+1,y)}^{(l)} & = & p+\alpha \left( \hat{e}_{x}\cdot 
\hat{u}\right) \sigma _{k}^{l}, \\ 
&  &  \\ 
P_{(x,y)\rightarrow (x-1,y)}^{(l)} & = & p+\alpha \left( -\hat{e}_{x}\cdot 
\hat{u}\right) \sigma _{k}^{l}, \\ 
&  &  \\ 
P_{(x,y)\rightarrow (x,y+1)}^{(l)} & = & p+\alpha \left( \hat{e}_{y}\cdot 
\hat{u}\right) \sigma _{k}^{l}, \\ 
&  &  \\ 
P_{(x,y)\rightarrow (x,y-1)}^{(l)} & = & p+\alpha \left( -\hat{e}_{y}\cdot 
\hat{u}\right) \sigma _{k}^{l},%
\end{array}
\label{move}
\end{equation}%
and it remains in its cell according to 
\begin{equation}
P_{(x,y)\rightarrow (x,y)}^{(l)}=1-\sum\limits_{\langle x^{\prime
},y^{\prime }\rangle }P_{(x,y)\rightarrow (x^{\prime },y^{\prime
})}^{(l)}=1-4p,  \label{stay}
\end{equation}%
where $p$ is a constant probability of moving to one of its first
neighboring cells with $0<p\leq 1/4$, $\alpha $ is a parameter that controls
the bias towards the bar with $0<\alpha \leq p$, $\hat{e}_{x}$ and $\hat{e}%
_{y}$ are the base vectors of our reference frame precisely as shown in Fig. %
\ref{Fig:night_club} (b), $\vec{u}=\vec{u}(\vec{r})$ is the floor field that
points to the nearest cell of the bar where $\hat{u}=\vec{u}/\left\Vert \vec{%
u}\right\Vert $, $\vec{r}=x\hat{e}_{x}+y\hat{e}_{y}$, with $x,y=1,...,L$, is
the agent position vector, and $\sigma ^{l}$ is the state of the agent and
assumes the values 
\begin{equation}
\sigma _{k}^{l}(\vec{r})=\left\{ 
\begin{array}{ccc}
1, & \text{if} & \text{state }(i), \\ 
-1, & \text{if} & \text{state }(ii), \\ 
0, & \text{if} & \text{if state }(iii),%
\end{array}%
\right.  \label{func}
\end{equation}%
according to the criteria stated in the sixth paragraph of this section. The
inner product inside the parenthesis of Eq. \ref{move} reflects the
contribution of the base vector pointing at the direction of the
corresponding transition to the probabilities of movement towards/away from
the bar direction.

In a nightclub venue, revenue usually comes primarily from sales of
alcoholic beverages at the bar. Depending on the event type, entry tickets
may also be a source of revenue, which we will also consider in our work.
However, we will first focus on the bar earnings. At a given time step $l$
of the venue, the number of attendees buying a drink at the bar
simultaneously, here denoted as $N_{d}(l)$, is lower or equal to the number
of the cells $N_{a}\ $adjacent to the bar, i.e., $N_{d}(l)\leq N_{a}$. Thus,
one defines the income at the bar per time step $l$ as: 
\begin{equation}
w^{(l)}=c_{d}N_{d}(l),  \label{rate}
\end{equation}%
which if we sum it over the time duration of the venue, it gives us the bar
accumulated earnings: 
\begin{equation}
W=\sum\limits_{l=0}^{t_{\max }}w^{(l)}=c_{d}\sum\limits_{l=0}^{t_{\max
}}N_{d}(l).  \label{Eq:acumulated}
\end{equation}

A more interesting amount is the adimensional one, obtained by dividing $W$
by $t_{\max }$, $c_{d}$, and $N_{a}$, which represents an earning rate of
bar: 
\begin{equation}
\overline{w}=\frac{W}{c_{d}\ N_{a}\ t_{\max }}=\frac{1}{\ \ t_{\max }N_{a}}%
\sum\limits_{l=0}^{t_{\max }}N_{d}(l).  \label{Eq:mean}
\end{equation}

To understand the essential aspects of the model, by focusing on this
quantity $\overline{w}$, it is crucial to consider the simplest scenario:
unbiased motion ($\alpha =0$), without region 2 in the lattice ($b=0$), and
without time interval between drinks ($\tau _{k}=0$) for every agent at
thermodynamic limit. In the mean-field regime, we expect that each cell can
be occupied, on average, by $\rho \equiv N/V$ particles once agents are
performing random walks with exclusion principles (standard lattice gas
dynamics). Similarly, on average, any given cell will be found "empty" with $%
1-\rho $ particles.

\begin{figure*}[t]
\begin{center}
\includegraphics[width=\textwidth]{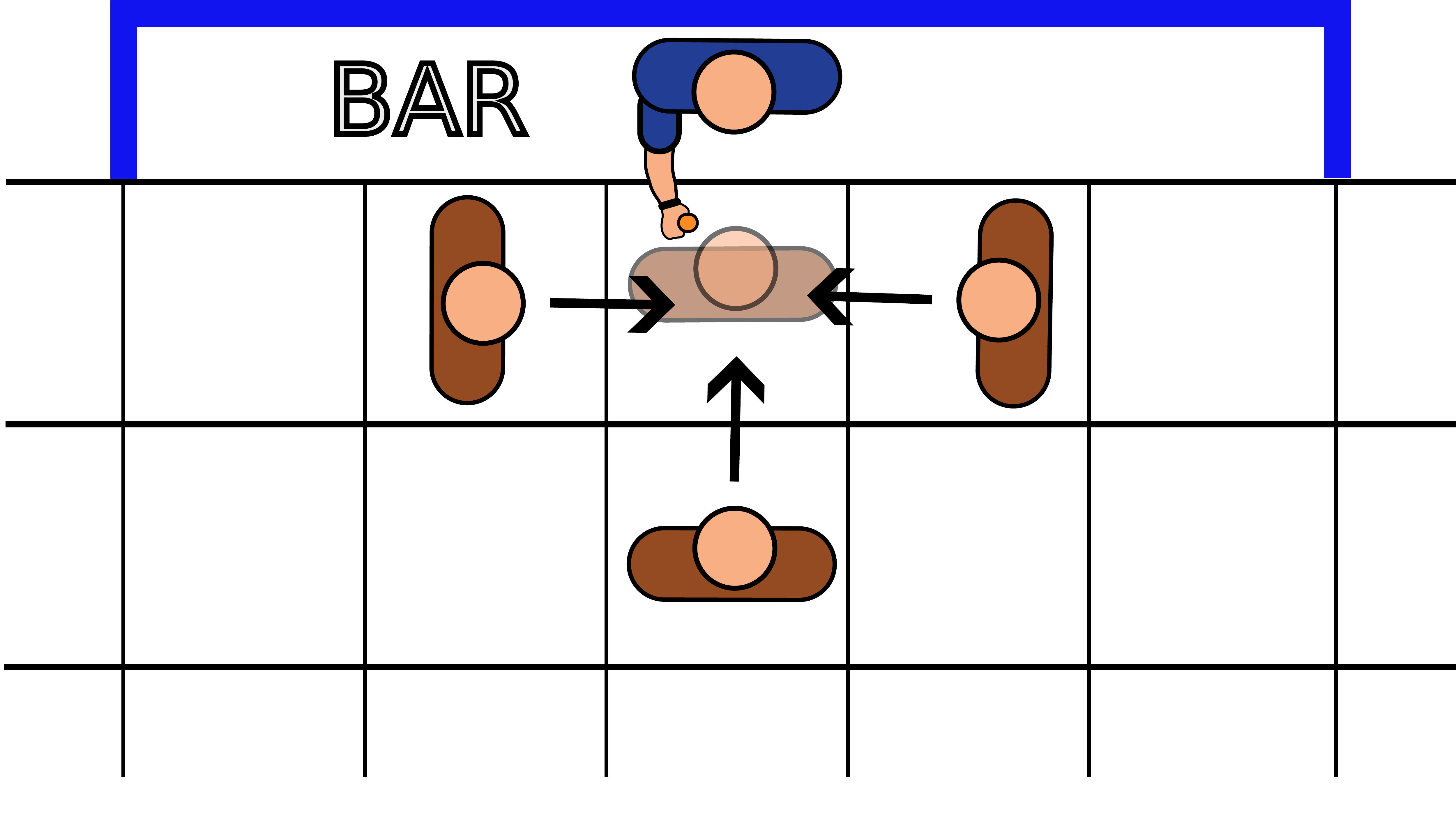}
\end{center}
\caption{ Possible movements that contribute to revenue to the bar. An agent
advances to an empty cell in the balcony or due to lateral movements.
Additionally, when the agent stays still on the counter also generates
revenue.}
\label{Fig:bar}
\end{figure*}

Each bar's target cell (empty) has only three neighboring cells from which
an agent may come: from its nearest left neighboring cell on the counter,
from its neighboring right position on the counter, i.e., lateral movements,
or finally from its nearest neighboring frontal position, in this case,
advancing to the bar counter as shown in Fig. \ref{Fig:bar}. 

In addition, another contribution to the bar's revenue occurs when an agent
on the balcony remains stopped in the same place. Thus considering these two
contributions, mean-field approximation results in the following: 
\begin{equation*}
\begin{array}{lll}
\left\langle N_{d}(l)\right\rangle & \approx & 3pN_{a}\rho (1-\rho
)+(1-4p)N_{a}\rho, \\ 
&  &  \\ 
& = & -3pN_{a}\rho ^{2}+(1-p)N_{a}\rho.%
\end{array}%
\end{equation*}

Thus, if we consider $p=1/4$, which is an appropriate choice in this case
(the probability of the agent remaining in the same cell is zero), the
mean-field approximation of the quantity represented by Eq. \ref{Eq:mean}
is: 
\begin{equation}
\begin{array}{lll}
\left\langle \overline{w}\right\rangle & = & \frac{1}{\ \ N_{a}\ t_{\max }}%
\sum\limits_{l=0}^{t_{\max }}\left\langle N_{d}(l)\right\rangle, \\ 
&  &  \\ 
& = & \frac{3}{4}\rho (1-\rho ).%
\end{array}
\label{Eq:profit_mean_field}
\end{equation}

We can observe in Eq. \ref{Eq:profit_mean_field} that maximum profits will
be achieved by the nightclub when $\rho =0.5$, i.e., the number of attendees
is half the capacity. This first result gives us a clue about how pedestrian
dynamics influence earnings with the sale of drinks.

With this simple approximation as reference, in the next section, we will
present our numerical results from the lattice gas model via MC simulations.
We used asynchronous updating scheme for the agent's position and we analyze
the deviations from the simple approximation described by Eq. \ref%
{Eq:profit_mean_field} by studying all the parameters and details not
explored in the simple mean-field approximation. 
%We also analyzed the accumulated earning time series and, thus, the earning rate time series, both averaged over a finite number of equally prepared samples (also denoted as runs in this manuscript). For simplicity, we consider that each attendee has an infinite amount of money and $p=1/4$ in all our simulations. 
We analyzed bias and memory effects on the distribution shape of the
nightclub earnings as a function of the density of agents. Finally, we
present an additional analysis focusing on profit maximization while looking
into the relation between ticket and drink prices. The idea of this study is
to work as a recommendation for bar owners to keep their profits whilst
keeping the attendees safe.

\section{Results}

\label{Sec:Results}

In this section, we present our results from numerical simulations. To
optimize the computational time, we correlate the number of runs, $N_{\text{%
runs}}$ related to the number of agents, $N$, considered on each case
according to the following relation: $N_{\text{runs}}=N_{\text{max-runs}%
}(1-\rho )+N_{\text{min-runs}}$, where $\rho =N/V=N/L^{2}$, $N_{\text{%
max-runs}}=10^{5}$, and $N_{\text{min-runs}}=10^{6}/L$.

We first show the results of the simplest case in which agents move randomly
through the lattice as in a standard lattice gas model. In that scenario, we
will only focus on the bar revenue and will consider $\alpha =0$ with all
agents having memory time $\tau =0$. We first simulate for venue duration of 
$t_{\max }=10^{4}$ MC steps, $b=0$ and $a=L$. Thus, we study the earning
rate given by Eq. \ref{Eq:mean} as a function of $\rho$.

\begin{figure*}[t]
\begin{center}
\includegraphics[width=\textwidth]{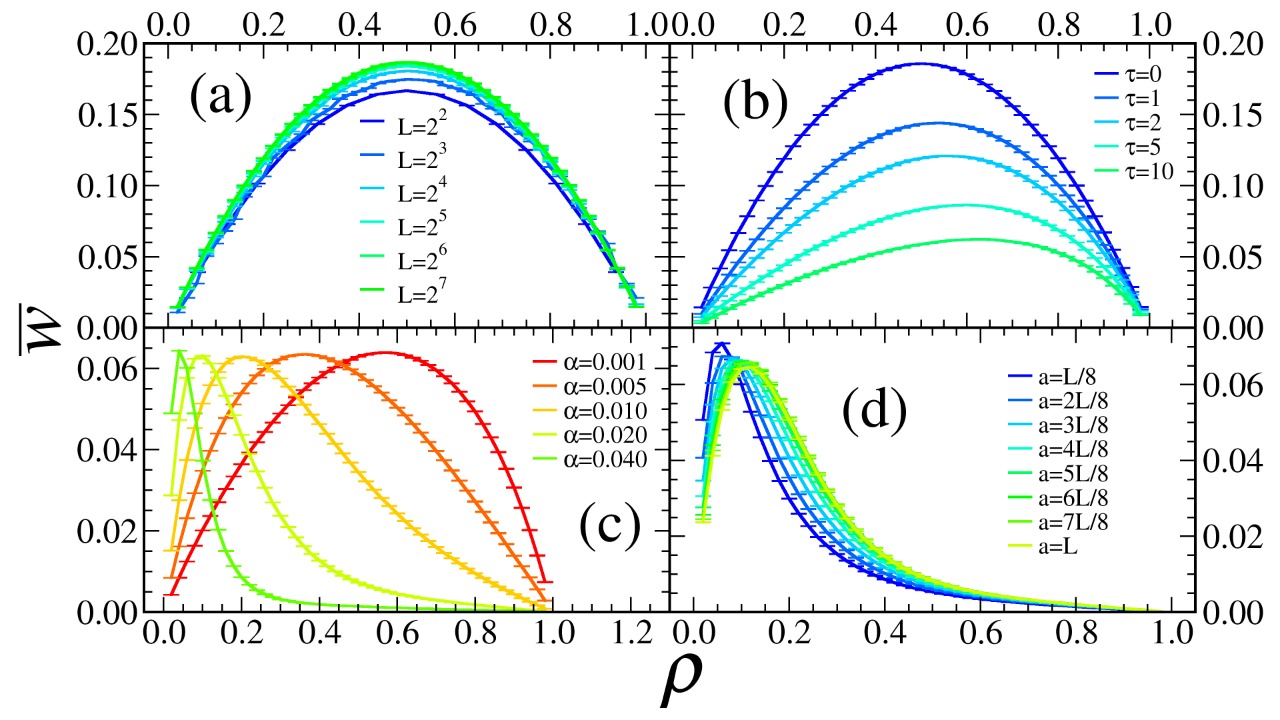}
\end{center}
\caption{Study of average earning rate after $t_{\max }=10^{4}$ MC steps
versus density for different intensive and extensive parameters of our
model. Plot (a) shows the finite size scaling for the simplest case. Plot
(b) shows the influence of different memory times between drinks for a
system with dimension $L=2^{6}$ (also used in plots (c) and (d)). Plot (c)
shows the influence of bias in the movement of agents inside the nightclub
for a homogeneous distribution of $\protect\tau =10$. Finally, plot (d)
shows the finite-size scaling of the bar's length.}
\label{Fig:basic}
\end{figure*}

Fig. \ref{Fig:basic} (a) shows a finite-size scaling analysis of the system.
We observe that $\overline{w}$ presents an inverted parabolic shape as a
function of the density of attendees (a fixed number for the whole venue
duration) for different lattice sizes. For $L$ sufficiently large ($L\approx
2^{6}$), we have a perfect match with the mean-field result (Eq. \ref%
{Eq:profit_mean_field}). It is essential to observe that $\overline{w}$
numerically arrives at the expected value $\left\langle \overline{w}%
\right\rangle _{\max }=\frac{3}{4}\frac{1}{2}\frac{1}{2}=\frac{3}{16}%
=\allowbreak 0.187\,5$ for $\rho =1/2$. Thus it is crucial to consider that
both simulations and mean-field corroborate the conclusion that for greater
than half the system capacity, in the simplest case, the total revenue
decreases due to overall jamming.

Now, alternatively, we studied the effects of non-zero memory, but
similarly, with all agents having the same memory value $\tau \neq 0$
(homogeneous distribution), things start to get interesting. From this
point, we studied only the system for an appropriate large system: $L=2^{6}$%
. Fig. \ref{Fig:basic} (b) shows the earning rate of the bar for different
values of $\tau $. We observe that the nightclub revenue rate decreases for
all values of density as attendees' time between drinks increases. We expect
such behavior when we remember that we are considering the same fixed party
duration and increasing the time for attendees to want another drink. The
non-trivial part is that the density in which the maximum revenue rate is
obtained shifts slightly to greater density values, indicating that a
greater number of attendees is required to achieve maximum profits. For
example, when $\tau =10$, the maximum profit is obtained at $\rho \approx 0.6
$.

We thus studied the influence of biased motion, i.e., $\alpha \neq 0$, for
homogeneous distribution with $\tau =10$. Fig. \ref{Fig:basic} (c) shows
that as $\alpha $ increases, the earning tax of bar curves starts to present
an increasing asymmetry in their shape. The density of maximum revenue rate
shifts to the left (inverse behavior compared with $\tau $) as $\alpha $
increases. However, the maximum revenue plateau is kept qualitatively
unchanged in this case. In this situation, one expects the stronger the
directed movement, the more likely is the occurrence of jamming at the bar
region for the same average density value. Even so, for smaller values of $%
\rho $, the maintenance of the plateau suggests that biased motion favors
consumption at the bar.

Finally, Fig. \ref{Fig:basic} (d) presents a finite size scaling of the bar
width in terms of $L$. In this case, we used $\alpha =0.020$ and observed
that the curves' peak shifts slightly to the right and downwards from $\rho
\approx 0.05$ to $\rho \approx 0.1$ as the bar size increases from $a=L/8$
to $a=L$. We observe no qualitative difference on $\overline{w}$ for $a\geq
3L/8$. 
\begin{figure}[t]
\centering%
\begin{subfigure}[t]{0.31\textwidth}
\centering
\includegraphics[width=\textwidth,trim={22cm 5cm 22cm 5cm},clip]{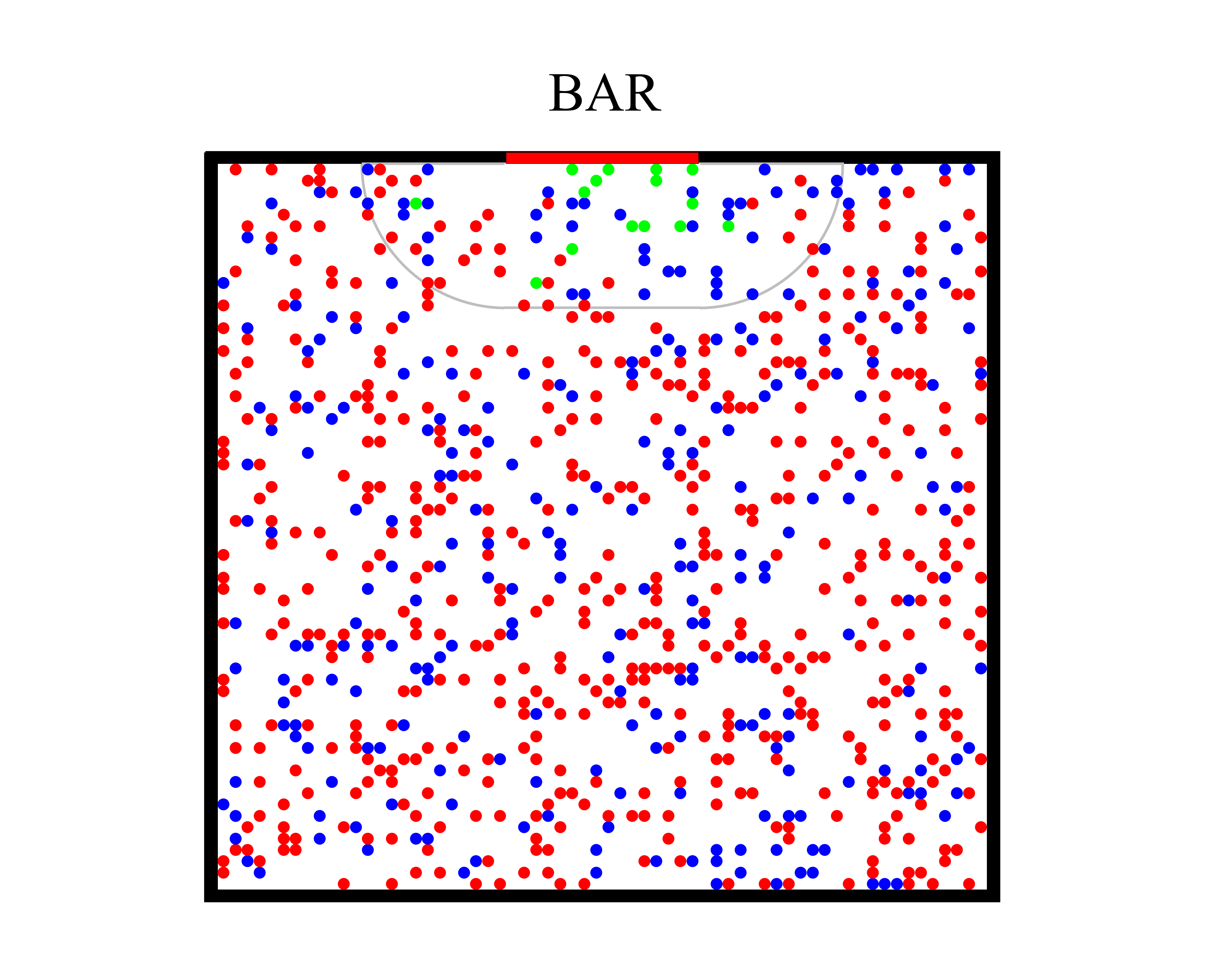} 
\caption{$\alpha=0.0$.} \label{fig:snap_4}
\end{subfigure}\hfill 
\begin{subfigure}[t]{0.31\textwidth}
\centering
\includegraphics[width=\textwidth,trim={22cm 5cm 22cm 5cm},clip]{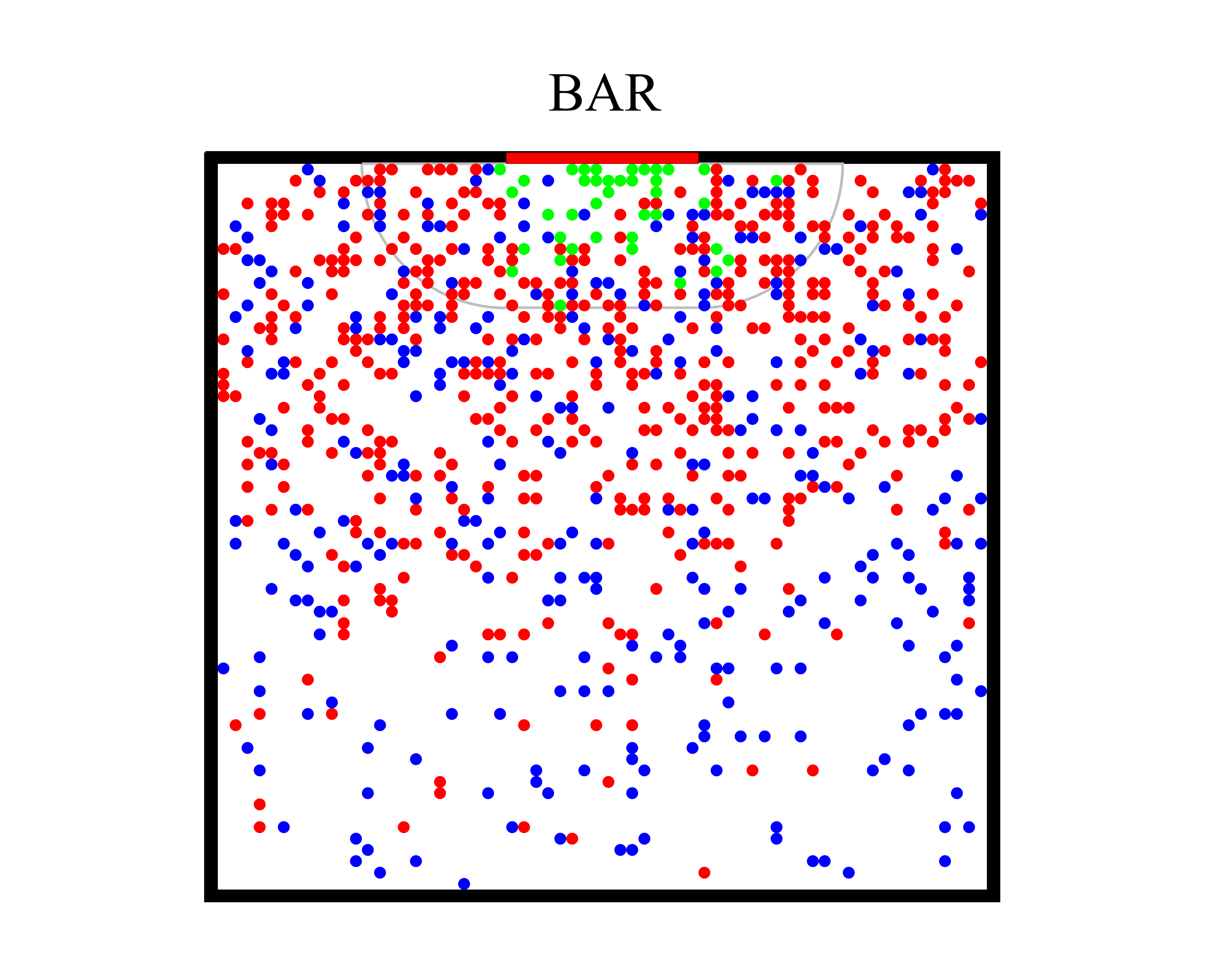} 
\caption{$\alpha=0.01$.} \label{fig:snap_5}
\end{subfigure}\hfill 
\begin{subfigure}[t]{0.31\textwidth}
\centering
\includegraphics[width=\textwidth,trim={22cm 5cm 22cm 5cm},clip]{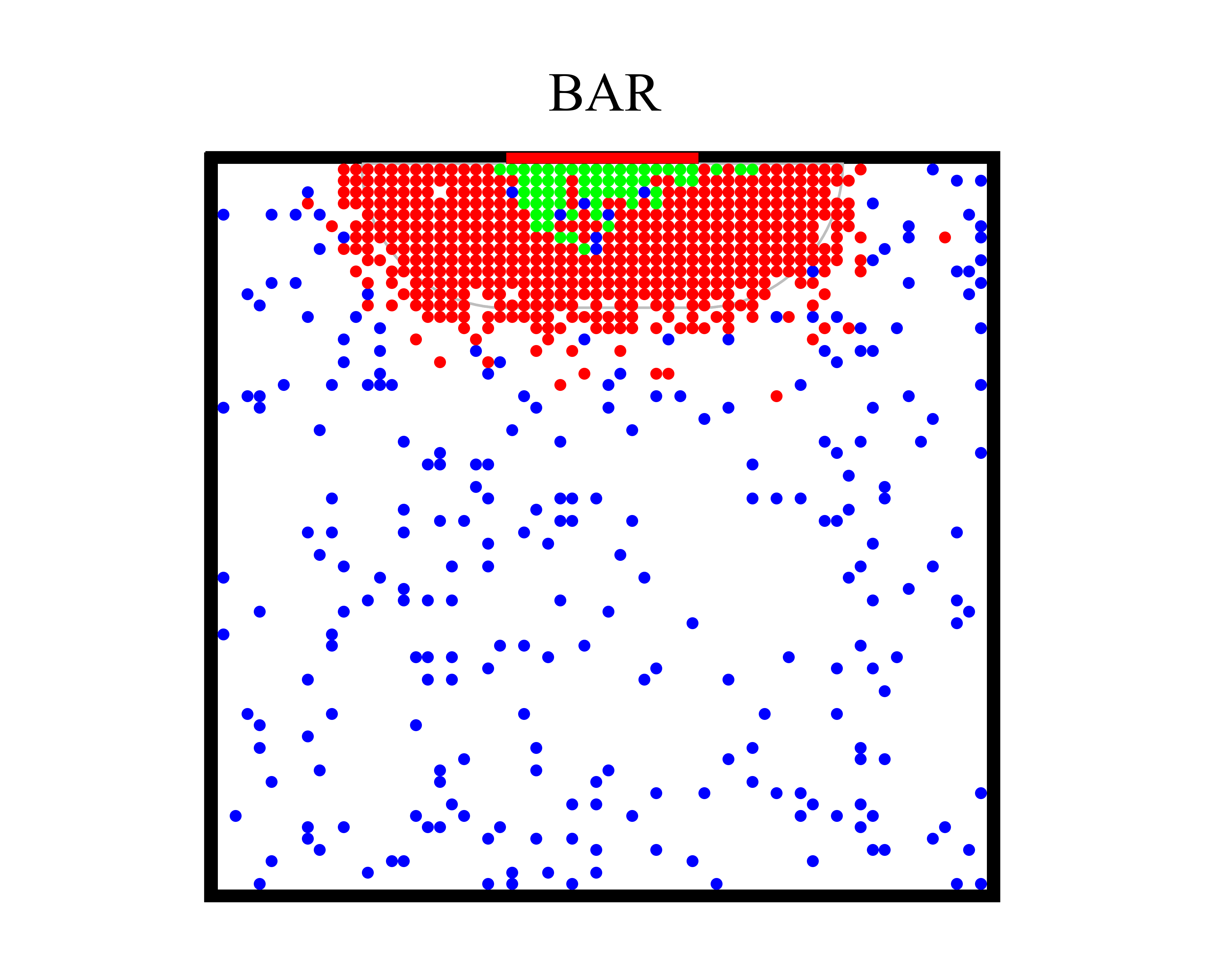} 
\caption{$\alpha=0.1$.} \label{fig:snap_6}
\end{subfigure}
\caption{Snapshots of the nightclub dance floor after $t=10^{4}$ MC steps
for different values of $\protect\alpha $. We used a power-law distribution
with exponent $\protect\gamma =1.1$ and minimum memory time of $\protect\tau %
=256$. We also used $N=800$, $L=64$, $a=2^{4}$ and $b=12$.}
\label{snapshots_alpha}
\end{figure}

We now introduce a more realistic ingredient, a non-homogeneous distribution
of $\tau $. We define this memory time by following a power-law distribution,

\begin{equation*}
P(\tau )\sim \tau ^{-\gamma }
\end{equation*}

In performing that, we can reproduce different sets of the population by
adjusting the exponent $\gamma $of the power law distribution by including
the anomalous situations: $\gamma <2$. Another more real aspect that we
consider in our analysis is the minimum memory time of $\tau =256$ so that
even the most compulsive drinker in the simulation has a reasonable time to
"walk" and "dance" through the lattice before rushing to the bar for another
drink.

To understand how this may affect the system, we show three snapshots of the
spatial distribution of agents in Figs. \ref{snapshots_alpha} and \ref%
{snapshots_gamma} after $t_{\max }=10^{4}$ MC steps of the simulation for
extensive parameters $N=800$, $L=2^{6}$, $a=2^{4}$, and $b=12$. In the
snapshots, agents in red are in state $(i)$, agents in green are in state $%
(ii)$, and agents in blue are in the state $(iii)$. The snapshots in Fig. %
\ref{snapshots_alpha} show the spatial distribution for different values of $%
\alpha $ using $\gamma =1.1$. In plot (a), we observe the case of non-biased
movement ($\alpha =0.0$). In that case, we observe that agents in states $(i)
$ and $(iii)$ are uniformly distributed spatially, except for the agents in
state $(ii)$ inside region two. Plot (b) of Fig. \ref{snapshots_alpha} shows
a mildly biased scenario, $\alpha =0.01$. In that case, we observe a higher
density of agents in state $(i)$ for agents near the bar. In plot (c) of
Fig. \ref{snapshots_alpha}, we observe that $\alpha =0.1$ corresponds to an
extreme scenario for the considered density, for which jamming occurs due to
agents presenting "zombie"-like behavior while trying to reach the bar. 
\begin{figure}[t]
\centering%
\begin{subfigure}[t]{0.31\textwidth}
\centering
\includegraphics[width=\textwidth,trim={22cm 5cm 22cm 5cm},clip]{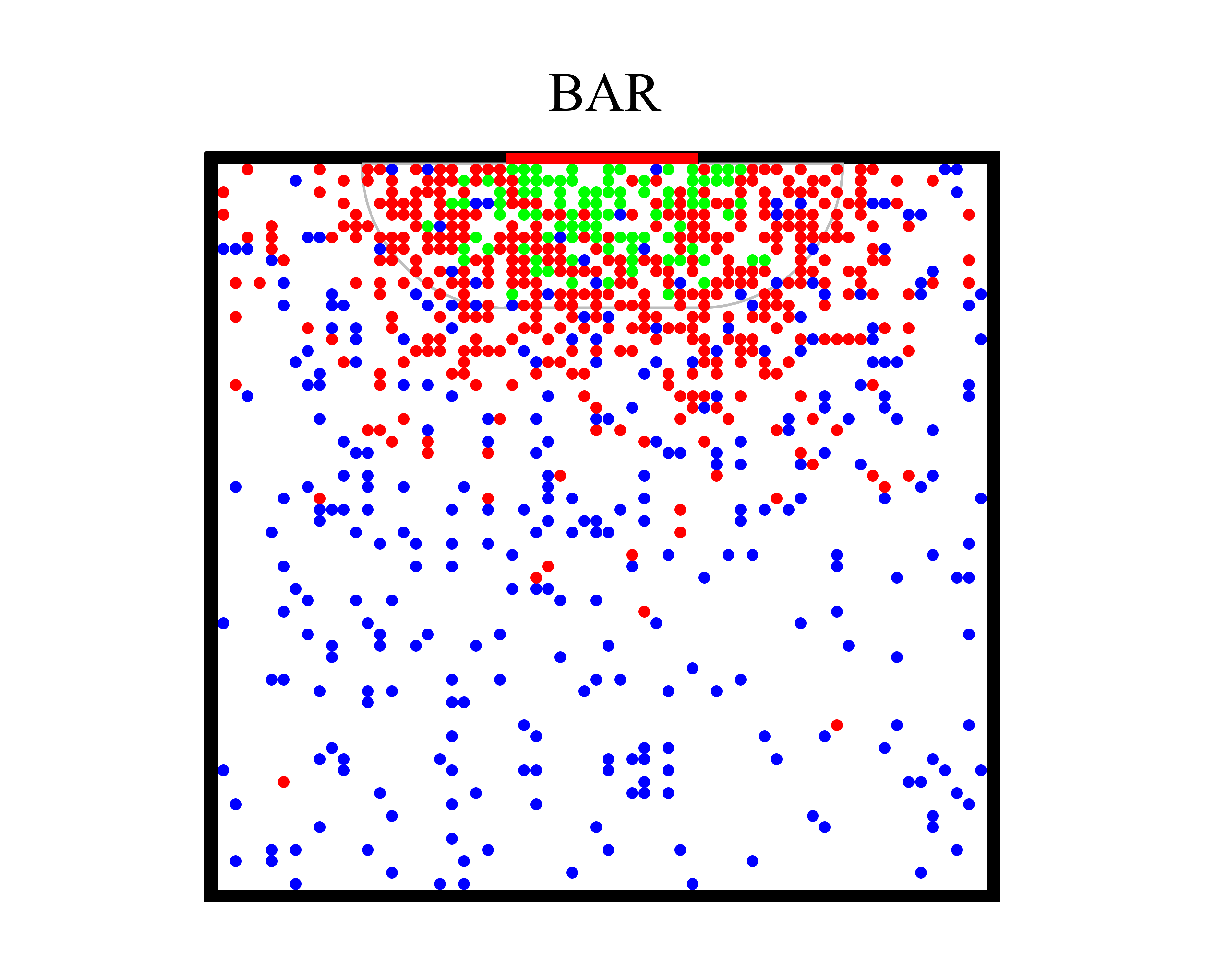} 
\caption{$\gamma=1.1$.} \label{fig:snap_1}
\end{subfigure}\hfill 
\begin{subfigure}[t]{0.31\textwidth}
\centering
\includegraphics[width=\textwidth,trim={22cm 5cm 22cm 5cm},clip]{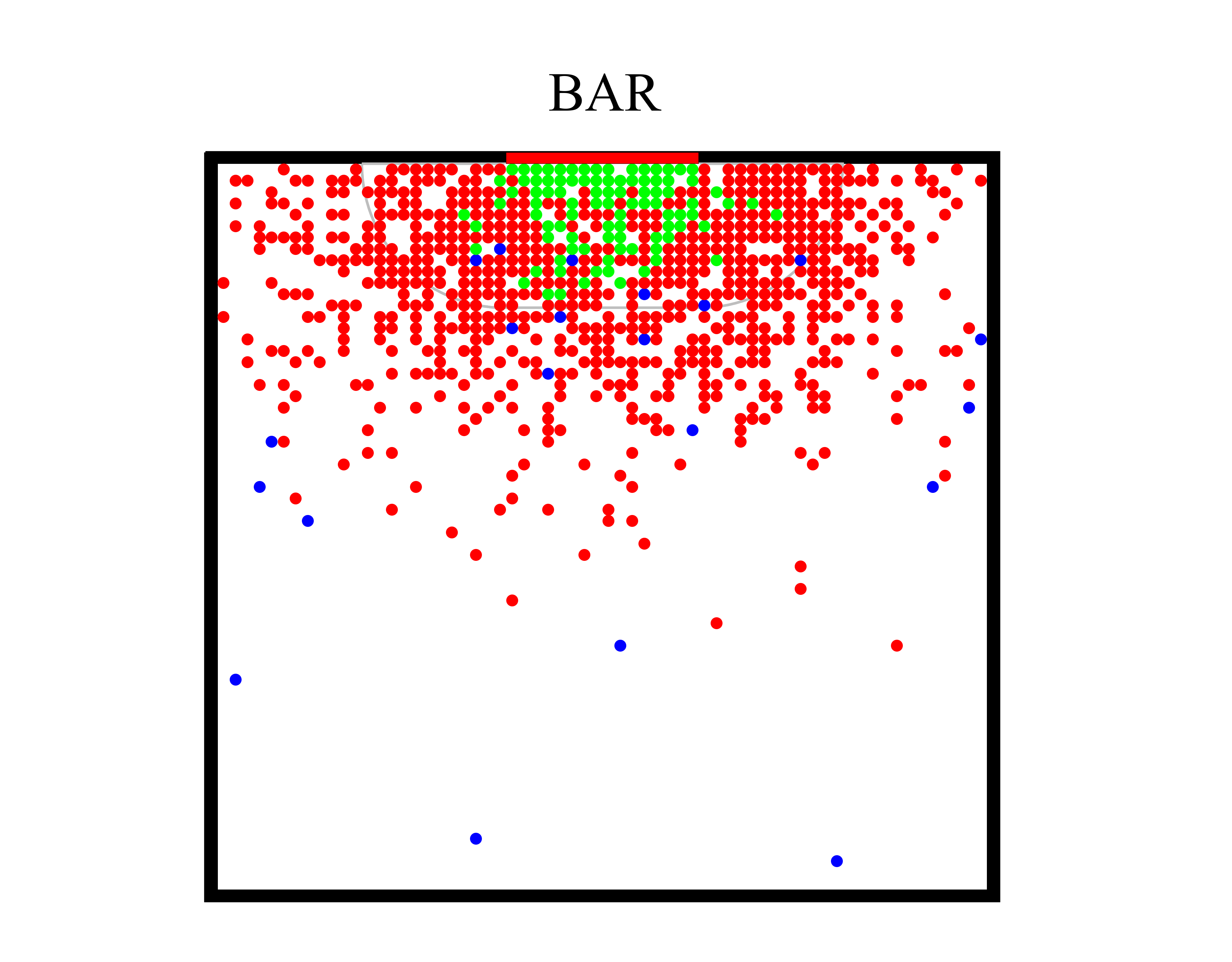} 
\caption{$\gamma=1.5$.} \label{fig:snap_2}
\end{subfigure}\hfill 
\begin{subfigure}[t]{0.31\textwidth}
\centering
\includegraphics[width=\textwidth,trim={22cm 5cm 22cm 5cm},clip]{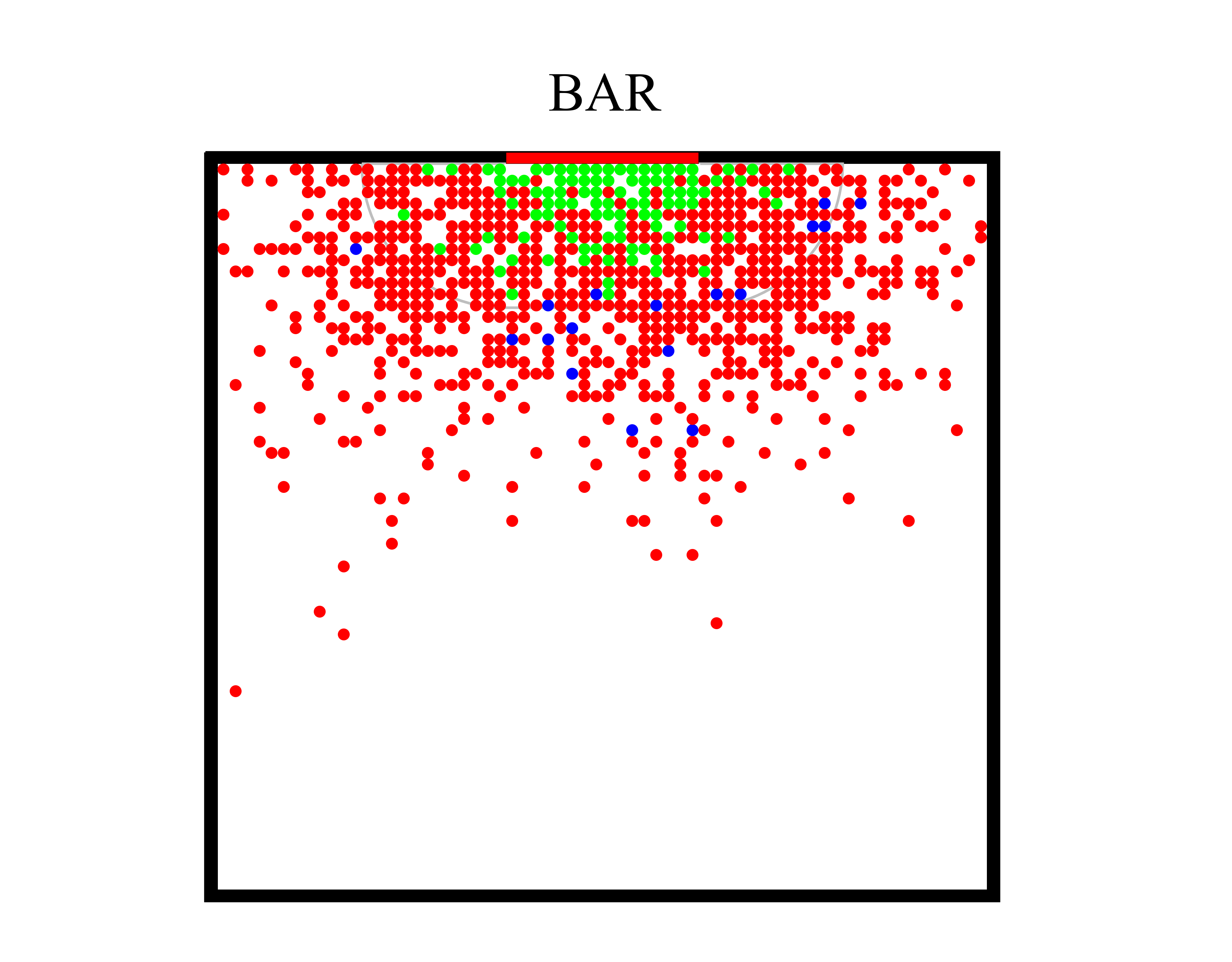} 
\caption{$\gamma=6.0$.} \label{fig:snap_3}
\end{subfigure}
\caption{Snapshots of the nightclub dance floor after $t=10^{4}$ MC steps
for different values of the power-law distribution exponent $\protect\gamma $%
. We used $N=800$, $\protect\alpha =0.03$, $L=64$, $a=2^{4}$, and $b=12$.}
\label{snapshots_gamma}
\end{figure}

Now we focus our attention on the distribution of memory time. In Fig. \ref%
{snapshots_gamma}, we show three snapshots for different values of power-law
exponent, $\gamma $, while keeping the same parameters as in Fig. \ref%
{snapshots_alpha}, except for the level of bias that we now used $\alpha
=0.03$. Here, we also kept the minimum value of memory time at $\tau =256$.
In plot (a) of Fig. \ref{snapshots_gamma}, we show the spatial distribution
of agents for $\gamma =1.1$. In this case, we observe a similar situation as
the previous case, but with an intermediate value of $\alpha $ in comparison
with Fig. \ref{snapshots_alpha} (a) and (b). In plot (b) of Fig. \ref%
{snapshots_gamma}, we show the spatial distribution for $\gamma =1.5$. We
observe that the number of agents in the state $(iii)$ after $t_{\max
}=10^{4}$ that are far from region two, which is in a pre-jammed state.
Agents in the state $(iii)$ far from region 2 are the ones to be considered
abstemious with this particular value of $\gamma $. In plot (c) of Fig. \ref%
{snapshots_gamma}, we show the snapshot for $\gamma =6.0$. In this
situation, we reproduce a homogeneous distribution with $\tau =256$ and
observe a similar behavior as plot (b). However, here agents in state $(iii)$
(blue) are only found near the boundary between regions 2 and 3, for it is
where the transition from the state $(ii)$ to state $(iii)$ is defined to
happen.

% \begin{figure}{L}{0.5\textwidth}
% \begin{center}
\begin{wrapfigure}{l}{0.5\textwidth}
\includegraphics[width=0.5\textwidth]{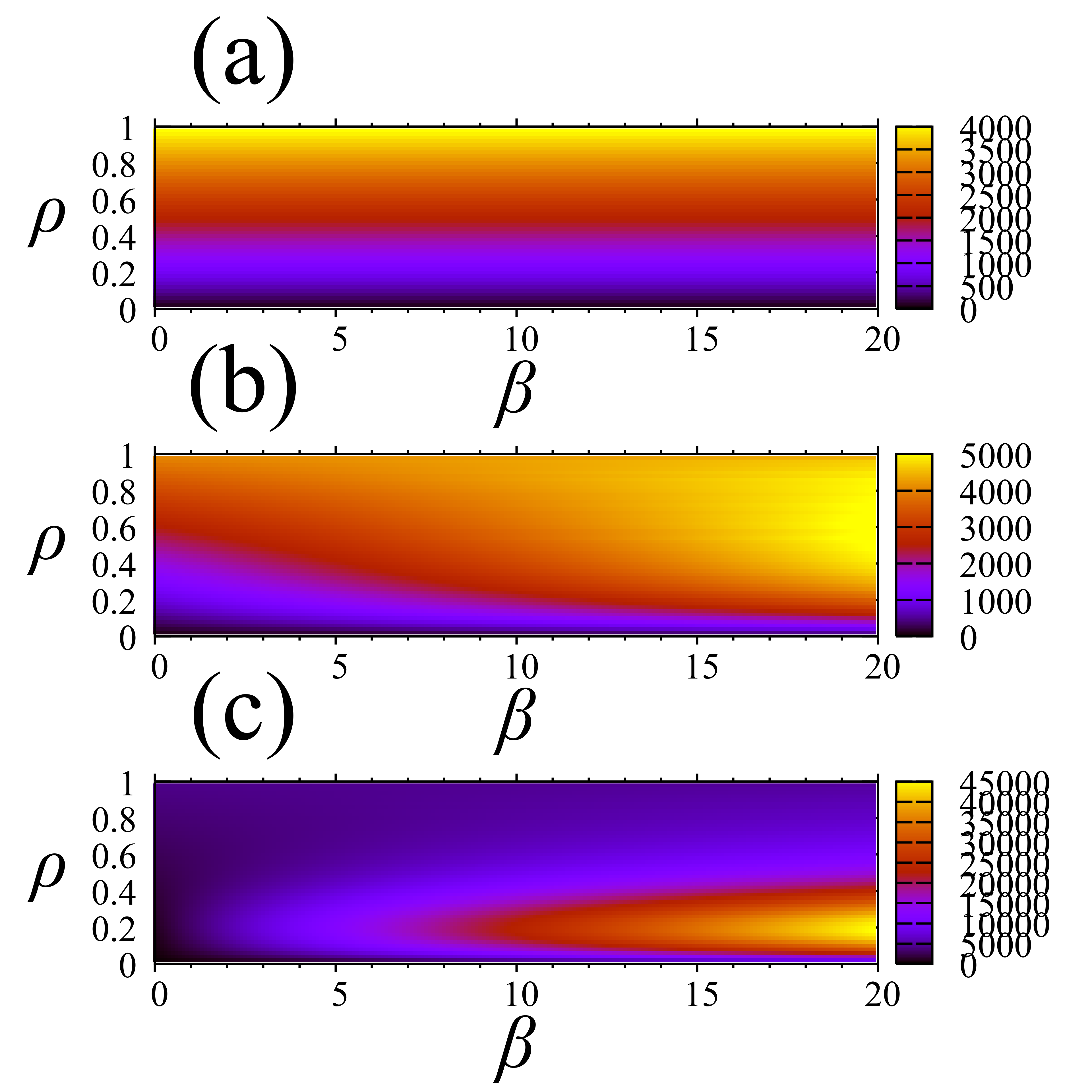}
% \end{center}
\caption{Map color of the average accumulated revenue as a function of density and the price ratio between drink price over ticket price for (a) $t=10^2$ MC steps, for (b) $t=10^3$, and for (c) for $t=10^4$ MC steps.}
\label{Fig:color}
% \end{figure}
\end{wrapfigure}

Now that we have established a basic understanding of how the intensive and
extensive parameters rule the dynamics and influence the accumulated gains
at the bar of our simulated nightclub, we shall focus our attention on the
relation between ticket and drink prices. We consider the accumulated amount:%
\begin{equation}
W_{total}=c_{d}\sum\limits_{l=0}^{t_{\max }}N_{d}(l)+Nc_{t}
\label{Eq:Wtotal}
\end{equation}%
where the first term represents the accumulated earnings, while the earnings
with entry tickets (second term) are $W_{t}=Nc_{t}$. To properly study both
sources of income, we define the ratio $\beta \equiv c_{d}/c_{t}$, where $%
c_{t}$ is the ticket cost. We studied this ratio in the range $0<\beta <20$,
while keeping $c_{t}=1$ fixed. We used $\alpha =0.02$, $\gamma =1.1$ with $%
\tau =256$ as minimal time, $L=2^{6}$, $a=2^{4}$, and $b=16$.

Fig. \ref{Fig:color}, we show the revenue accumulated over three different
event duration times as a function of the density and $\beta $. In plot (a)
of Fig. \ref{Fig:color}, we show the results for $t=10^{2}$ MC steps. In
this case, we observe that total revenue increases linearly with $\rho $,
once the simulation time is shorter than the minimal memory time, which
means that no consumption at the bar has happened yet. For $t=10^{3}$ shown
in plot (b), we notice that for small values of $\beta $, one obtains the
maximum earnings for high densities. However, we see that for $\beta \gtrsim
15$ the maximum revenue is obtained in an intermediate average density of
attendees because ticket and drink prices seem to be in a certain balance.
Finally, in plot (c) we show the revenue accumulated over $t=10^{4}$ MC
steps. We observe that for values in the range $0<\rho \lesssim 5$, the
accumulated revenue seems to present the same dependency on $\rho $ as we
saw in Fig. \ref{Fig:basic} (c), for instance. That means that, for drink
prices relatively greater than ticket prices, or even no entry ticket costs,
one does not obtain the maximum revenue at the total capacity of the
nightclub, not even close to it.

\section*{Summaries, conclusions, and discussions}

\label{Sec:Conclcusions}

Our model has shown that revenue from the sale of drinks in a nightclub is
highly sensitive to the density of attendees while describing the dynamics
with biased random walk rules directed by a static floor field. Our approach
captured the opposite influence that memory time between drinks and biased
movement had over the total revenue of drink sales.

We have shown that when entry tickets are part of the total revenue, it
becomes the primary source of income if the event duration is smaller than
the minimal memory time of attendees. However, for an intermediate event
duration time, we observe a balance between the gains at the bar and with
tickets when drinks prices are at least five times the ticket cost. For a
sufficiently long event duration, the drink sales become the primary source
of income of the nightclub, even if the drink prices are approximately equal
to ticket prices.

The model we proposed is quite simple and could answer the question proposed
at the end of the introduction: "Is it in the interest of a nightclub owner
to have its club at full capacity?". Furthermore, the answer is: "It
depends." If a nightclub owner has its primary source of income based on
ticket sales, then its maximum gains will occur with the nightclub at total
capacity. If there is another source of income inside the nightclub, like
drink sales, as we used here, then the maximum revenue is obtained from a
relatively low density of attendees depending on parameters that rule the
pedestrian dynamics of the event. Nightclub owners can control the average
density inside the nightclub by controlling entry queues or even with
detailed market research to establish drink and ticket prices. To add
complexity to our model, we can implement our model with a variable number
of attendees inside the nightclub by defining entering and exiting rates of
attendees, as well as considering a finite amount of money for each

% eal nightclub party does not have an
% infinite time of duration and the number of attendees fluctuates along a
% party's duration. However, for simplicity we also assume that each nightclub
% venue has a fixed number of attendees during its finite duration. That could be
% analogous to a nightclub party that all attendees arrive together at the
% beginning and leave together at end of the party.

% A more realistic approach would certainly include money restrictions, such as, for example, agent's money distribution according to a Pareto's law, but we will keep simple in this first study.

\section*{CRediT authorship contribution statement}

Eduardo Velasco Stock: Conceived and designed the analysis, Data curation,
Formal analysis, Wrote the paper, Elaborated the algorithms. Roberto da
Silva: Conceived and designed the analysis, Data curation, Formal analysis,
Wrote the paper, Elaborated the algorithms. HMust include all authors,
identified by initials, for example: A.A. conceived the experiment(s), A.A.
and B.A. conducted the experiment(s), C.A. and D.A. analysed the results.
All authors reviewed the manuscript.

\section*{Declaration of competing interest}

The authors declare that they have no known competing financial interests or
personal relationships that could have appeared to influence the work
reported in this paper

\section*{Acknowledgements}

The authors thank CNPq for financial support under grant number
311236/2018-9.

\bibliographystyle{elsarticle-num}
\bibliography{Bibliografia.bib}

\end{document}